\documentclass[aps,twocolumn,nofootinbib,preprintnumbers,superscriptaddress]{revtex4-1}
\pdfoutput=1

\usepackage[usenames,dvipsnames,table]{xcolor}
\usepackage{graphicx,amsmath,amssymb,amsthm,multirow,array,bm,}
\usepackage[mathscr]{eucal}
\usepackage[bbgreekl]{mathbbol}
\usepackage{amsfonts}
\usepackage{hyperref}
\usepackage{tikz}
\usetikzlibrary{3d,calc}
\usepackage{ulem}

\hypersetup{
    pdfstartview={FitH},    % fits the width of the page to the window
    pdftitle={Entanglement density and gravitational thermodynamics},    % title
    pdfauthor={Jyotirmoy Bhattacharya, Veronika Hubeny, Mukund Rangamani, Tadashi Takayanagi},     % author
    colorlinks=true,       % false: boxed links; true: colored links
    linkcolor=blue,          % color of internal links (change box color with linkbordercolor)
    citecolor=red,        % color of links to bibliography
    filecolor=magenta,      % color of file links
    urlcolor=blue           % color of external links
}
\definecolor{rust}{rgb}{0.8,0.2,0.2}

\def\Tr{{\rm Tr}}

\def\bdy{{\cal B}}

\def\regA{{\cal A}}

\def\rhoA{{\rho_{\regA}}}

\def\entsurf{
\partial \regA}

\def\domd#1{D[#1]}
\def\extr{{\cal E}_\regA}

\def\eden{\hat{n}}
\def\lcgen{{\mathscr C}}
\def\bme{{\bm \epsilon}}
\def\sa{{\sf a}}
\def\sc{{\sf c}}
\def\og{{\scriptscriptstyle \nearrow}}
\def\ig{{\scriptscriptstyle \nwarrow}}
\def\uL{u_{_L}}
\def\uR{u_{_R}}
\def\vL{v_{_L}}
\def\vR{v_{_R}}
\def\um{U}
\def\vm{V}
\def\ud{u_{_\delta}}
\def\vd{v_{_\delta}}

\newcommand{\cone}{
\begin{tikzpicture}
    \draw (0,0) arc (180:360:1mm and 0.5mm) -- (0.1,-0.2) -- cycle;
    \draw[dashed] (0,0) arc (180:0:1mm and 0.5mm);
    \shade[left color=blue!5!white,right color=blue!40!white,opacity=0.3] (0,0) arc (180:360:1mm and 0.5mm) -- (0.1,-0.2) -- cycle;
\end{tikzpicture}
}

\newcommand{\de}{\partial}
\newcommand{\be}{\begin{equation}}
\newcommand{\ba}{\begin{eqnarray}}
\newcommand{\ea}{\end{eqnarray}}
\newcommand{\ee}{\end{equation}}

\newcommand{\f}{\frac}
\newcommand{\s}{\sqrt}

\begin{document}

\title
{ Entanglement density and gravitational thermodynamics
}
\preprint{DCPT-14/77}
\preprint{YITP-104}
\preprint{IPMU14-0359}

\author{Jyotirmoy Bhattacharya}
\email{jyotirmoy.bhattacharya@durham.ac.uk}
\affiliation{Centre for Particle Theory \& Department of
Mathematical Sciences, Durham University, South Road, Durham DH1 3LE, United Kingdom}

\author{Veronika E. Hubeny}
\email{veronika.hubeny@durham.ac.uk}
\affiliation{Centre for Particle Theory \& Department of
Mathematical Sciences, Durham University, South Road, Durham DH1 3LE, United Kingdom}

\author{Mukund Rangamani}
\email{mukund.rangamani@durham.ac.uk}
\affiliation{Centre for Particle Theory \& Department of
Mathematical Sciences, Durham University, South Road, Durham DH1 3LE, United Kingdom}

\author{Tadashi Takayanagi}
\email{takayana@yukawa.kyoto-u.ac.jp}
\affiliation{Yukawa Institute for Theoretical Physics (YITP), Kyoto University, Kyoto 606-8502,
Japan}
\affiliation{Kavli Institute for the Physics and Mathematics of the Universe (WPI), University of Tokyo, Kashiwa, Chiba 277-8582, Japan}

%%%%%%%%%%%%%%%%%%%%%%%%%%%%%%%%%%%%%%%%%%%
\begin{abstract}
%%%%%%%%%%%%%%%%%%%%%%%%%%%%%%%%%%%%%%%%%%%
In an attempt to find a quasi-local measure of quantum entanglement, we introduce the concept of entanglement density in relativistic quantum theories. This density is defined in terms of infinitesimal variations of the region whose entanglement we monitor, and in certain cases can be mapped to the  variations of the generating points of the associated domain of dependence.
We argue that strong sub-additivity constrains the entanglement density to be positive semi-definite. Examining this density in the holographic context, we map its positivity to a statement of integrated null energy condition in the gravity dual. We further speculate that this may be mapped to a statement analogous to the second law of black hole thermodynamics, for the extremal surface.
\end{abstract}


\maketitle

%~~~~~~~~~~~~~~~~~~~~~~~~~~~~~~~~~~~~~~~~~~~~~~~
\section{Introduction}
\label{sec:intro}
%~~~~~~~~~~~~~~~~~~~~~~~~~~~~~~~~~~~~~~~~~~~~~~

The holographic AdS/CFT correspondence indicates that the fundamental constituents of spacetime geometry are quanta of a conventional non-gravitational field theory.  The precise manner in which these non-gravitational quanta conspire to construct  a smooth semiclassical spacetime, however, still remains obscure.
Holography is motivated by black hole thermodynamics, which suggests that emergence of gravity can be associated with coarse-graining a la classical thermodynamics \cite{Jacobson:1995ab}.  We then seek to understand what is being coarse-grained, and how.

A crucial hint is provided by the fact that AdS/CFT geometrizes quantum entanglement: entanglement entropy (EE) in the CFT is given by the area of a certain extremal surface in the bulk \cite{Ryu:2006ef,Ryu:2006bv,Hubeny:2007xt}. Indeed, the fascinating idea of spacetime geometry being the encoder of the entanglement structure of the quantum state \cite{Swingle:2009bg,VanRaamsdonk:2009ar,Maldacena:2013xja} hints at potentially deep insights into the workings of quantum gravity.

As a first step, we would like to decipher the dynamical equations of gravity from the these statements. In this regard, EE which motivates the connection to geometry, a-priori presents a complication: it is non-local -- even in local QFTs, it is defined on a causal domain.  The corresponding bulk quantity depends on the bulk geometry along a codimension-2 extremal surface. To make contact with local gravitational physics, it would be convenient to work with a more localizable construct in the dual CFT.\footnote{For recent progress on directly deriving gravitational dynamics from EE,  cf.,
\cite{Nozaki:2013vta,Bhattacharya:2013bna,Lashkari:2013koa,Faulkner:2013ica}.}

% Figure
\begin{figure}[htbp]
\begin{center}
\includegraphics[width=2in]{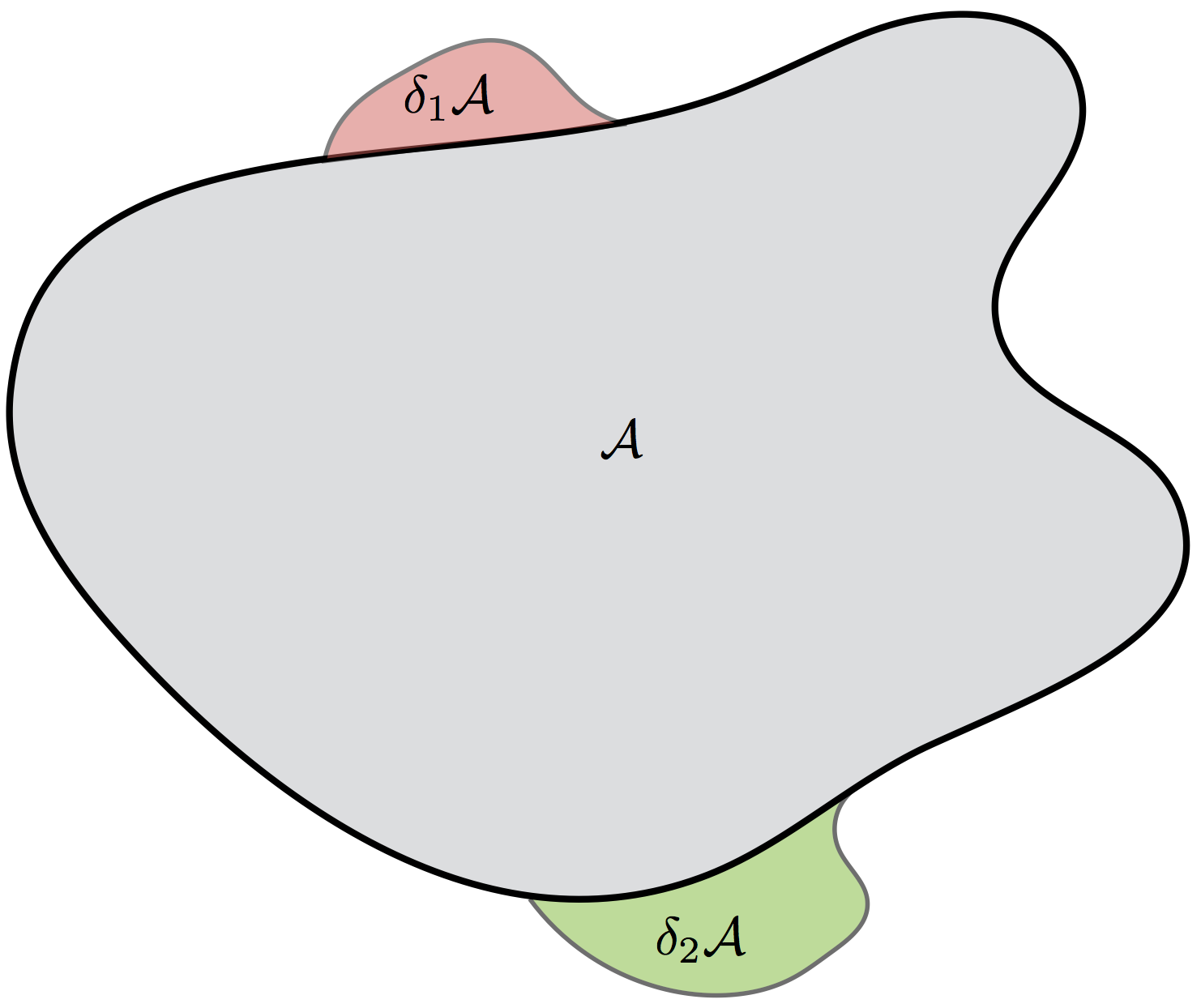}
\caption{Illustration of the generic variations $\delta_1 \regA$ and $\delta_2\regA$ which are used to define the entanglement density \eqref{eq:ndef1}. }
\label{fig:ndefine}
\end{center}
\end{figure}

Inspired by this logic, we propose to study  a  QFT quantity we call {\em entanglement density}.
This effectively measures two-body quantum entanglement between two infinitesimally small regions. To motivate its construction, consider a quantum field theory on a (rigid) background spacetime $\bdy$ which is foliated by spacelike Cauchy surfaces $\Sigma$. We pick a region $\regA \subset \Sigma$ and construct the reduced density matrix $\rhoA$.
The entanglement entropy
 $S_\regA = -\Tr\left(\rhoA \log \rhoA\right)$ is the von Neumann entropy of this density matrix, and is a functional of $\entsurf$.
We propose to retain locality by examining  EE for infinitesimal variations of  $\partial \regA$ (and hence $\regA$). Schematically for a configuration $\rho_\Sigma$ on the Cauchy slice, we define the double variation:\footnote{This construction has some parallels with recent discussions of differential entropy introduced in \cite{Balasubramanian:2013lsa} and explored more thoroughly \cite{Headrick:2014eia}.}
\begin{equation}
\eden\left(\delta_1 \regA, \delta_2 \regA \right)  = \delta_1\, \delta_2  \, S_\regA
\label{eq:ndef1}
\end{equation}	
The construction is pictorially illustrated in Fig.~\ref{fig:ndefine}.

Let us now simplify $\eden$ by appealing to the fact that $S_\regA$ is a functional on the entire domain of dependence
$\domd{\regA}$. We focus on backgrounds $\bdy$ and regions $\regA$ for which $\domd{\regA}$ is given by the intersection of past and future light-cones from two points, $\lcgen^\pm$ respectively. As a consequence we will focus on the variations inherent in \eqref{eq:ndef1} which are due to the variations of one of the points, say $\lcgen^-$, keeping the other fixed (or vice versa).\footnote{A related version of entanglement density was considered earlier in \cite{Nozaki:2013wia,Nozaki:2013vta}, without invoking the relativistic causal structure.} In this context $\delta_1\, \delta_2 \, S_\regA^{vac}=0$ for 2d and 3d CFTs, although  (\ref{eq:ndef1}) pertains in any QFT.

We will exploit the fact that $\eden$ is naturally sensitive to a key property of the von Neumann entropy, namely strong subadditivity (SSA), which states that
\begin{equation}
 S_{\regA_1 \cup \regA_2} + S_{\regA_1 \cap \regA_2} \leq S_{\regA_1} + S_{\regA_2}   \qquad \forall \,\regA_{1,2}\,.
\label{eq:ssa}
\end{equation}	
SSA is a convexity property of entanglement; for regions  in \eqref{eq:ssa} being small  deformations of a parent region, this has a  quadratic structure, which motivates \eqref{eq:ndef1}. Inspired by a beautiful construction of Casini \& Huerta \cite{Casini:2004bw,Casini:2012ei}, we show that entanglement density can be expressed as a second order  differential operator ${\cal D}_\pm^2$
acting on EE by differentiating with respect to the coordinates $\sc^\pm$ of $\lcgen^\pm$ (specified explicitly for $d=2,3$ in
\S\ref{sec:eessa}). SSA then implies ${\cal D}^2_{\sc^\pm} S(\sc^+;\sc^-) \geq 0$.

  Exploiting the holographic construction of EE in terms of  bulk codimension-two extremal surfaces $\extr$, we argue that the variations of interest can be mapped to the motion of the extremal surface along its  null normals $N_{(\pm)}^\mu$. Using standard differential geometric identities, this in turn can be simplified to a statement about the geometry side of Einstein's equations $E_{\mu\nu}= R_{\mu\nu}-\f{1}{2}Rg_{\mu\nu}+\Lambda \,g_{\mu\nu}$, namely
\begin{equation}
\int_{\extr}\, \bme \;  E_{\mu\nu} \, N_{(\pm)}^\mu\, N_{(\pm)}^\nu \geq 0 \,,
\label{eq:enngr}
\end{equation}	
where $\bme$ is the volume form induced on the extremal surface.\footnote{A sufficient condition for this positivity is the null energy condition. The null energy condition has been
crucial in the derivations of SSA \cite{Allais:2011ys,Callan:2012ip,Wall:2012uf}.} Indeed, as the main result of this paper we will show that the entanglement density is precisely given by (\ref{eq:enngr}) for small perturbations in the AdS$_3/$CFT$_2$ setup. We have therefore related SSA (which can be regarded as a physicality condition on EE) to a restriction on the spacetime curvature.\footnote{For other applications of entropic inequalities and related constraints in gravity duals see \cite{Blanco:2013joa,Banerjee:2014oaa,Bousso:2014uxa,Lin:2014hva}.}

{\em NB:} As this work was nearing completion we received \cite{Lashkari:2014kda}, where a similar relation between SSA and bulk energy stress tensor has been discussed. Similar results have been obtained by Arias and Casini (unpublished).

%~~~~~~~~~~~~~~~~~~~~~~~~~~~~~~~~~~~~~~~~~~~~~~~
\section{SSA in field theory}
\label{sec:eessa}
%~~~~~~~~~~~~~~~~~~~~~~~~~~~~~~~~~~~~~~~~~~~~~~

To set the stage for our analysis let us recall the proof of the c-theorem \cite{Zamolodchikov:1986gt} and F-theorem \cite{Myers:2010xs,Myers:2010tj,Jafferis:2011zi} based on SSA, as in \cite{Casini:2004bw,Casini:2012ei}. We consider subsystems which are defined by  the intersection of  light-cones
from two points $\lcgen^\pm$ in $d$-dimensional QFTs. Letting $\domd{\regA} = J^-[\lcgen^+] \cap J^+[\lcgen^-] $, we pick $\regA$ to be a Cauchy slice for $\domd{\regA}$ at constant time; see e.g., Figs~\ref{fig:2dSSA2} and \ref{fig:3dSSA}. Then $S_\regA$ can be viewed of as a function of the coordinates $\sc^\pm$ of $\lcgen^\pm$; i.e., $S_\regA \equiv S(\sc^+;\sc^-)$.
For $\bdy = {\mathbb R}^{d-1,1}$ we take $ \sc^\pm =( t_\pm, {\bf x}_\pm)$. Letting $\sa =\pm$, we define the entanglement density in $d=2,3$ with respect to varying $\lcgen^\pm$ as
\begin{equation}
\eden_\sa(t_\sa,{\bf x}_\sa) \equiv  \left[\Box_\sa\,  + \frac{2\,(d-2)}{t_\sa} \partial_{t_\sa} \right]S(t_\sa,{\bf x}_\sa)  \geq 0\,,
\label{eq:edeneqn}
\end{equation}
where the inequality is guaranteed by SSA.  We give a quick overview  following \cite{Casini:2012ei}, with some additional generalizations.
%
%----------------------------------------------------------------------
\subsection{QFTs in $d=2$}\label{ssec:CH1p1}
\begin{figure}[htbp]
\begin{center}
\includegraphics[width=2.3in]{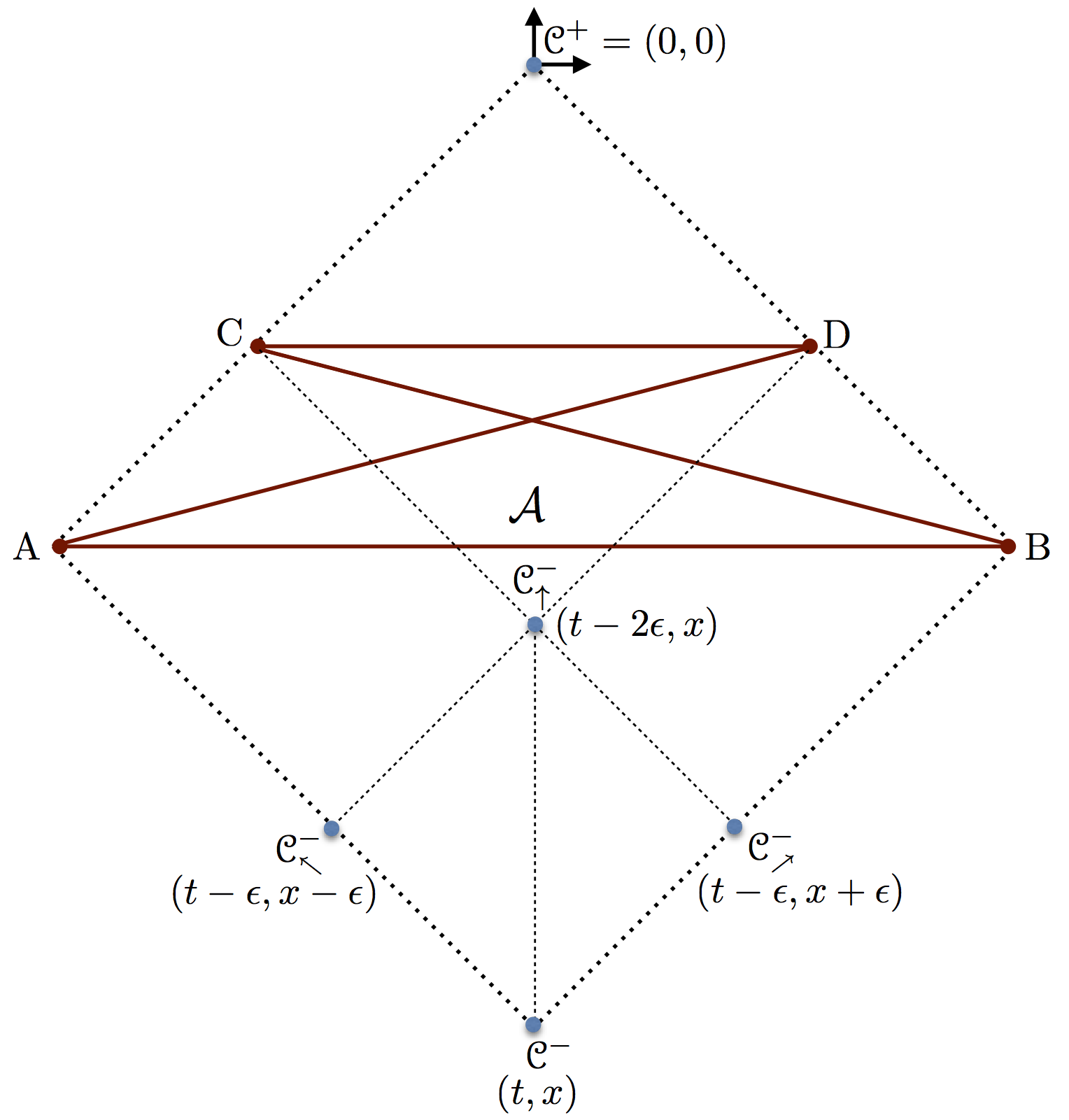}
\caption{Illustration of the set-up following \cite{Casini:2004bw} in $d=2$.  We choose $\lcgen^+$ to be the origin and the region $\regA$ lies on the time-slice with coordinate $\frac{1}{2}\, t$. We assume $t<0$ and $\epsilon \leq 0$. }
\label{fig:2dSSA2}
\end{center}
\end{figure}

%----------------------------------------------------------------------
%
We start by applying SSA to the configuration in Fig.~\ref{fig:2dSSA2}; for space- and time-translation invariant configurations, we can
w.l.o.g.\ fix $\lcgen^+ = (0,0)$ as a reference and drop subscripts for coordinates of $\lcgen^-$. SSA implies
\begin{equation}
 S_{AD}+S_{CB} \geq S_{AB}+S_{CD} \,.
 \label{2dssa1}
\end{equation}
The fact that EE is defined on a causal domain can be used to redefine our region. For example  $S_{AD} = S_{AC \;\cup \; CD}$
even for states which are not boost invariant,\footnote{Since we have null segments, this statement should be viewed in a suitable limiting sense.} since  both $AD$ and $AC\,\cup \,CD$ have the same domain of dependence. As a result we do not make any symmetry assumptions about the state for which EE is evaluated.

Now consider moving $\lcgen^-$ from its original location $(t,x)$ along the light-cone directions to $\lcgen^-_\og$
and $\lcgen^-_\ig$ respectively by an amount $\epsilon$. This effectively shifts the left and right end-points of $\regA$  along the boundary of $\domd{\regA}$ defining the regions on the l.h.s.\  of \eqref{2dssa1}. For the second region on the r.h.s.\ we can equivalently consider translating $\lcgen^- \mapsto \lcgen^+_\uparrow$ by a distance $2\epsilon$.
 Under these shifts we track the implications of SSA \eqref{2dssa1}. In fact, in the present case we simply need to plug in the explicit dependence of the coordinates of the end-points of the various regions:
\begin{equation}\label{alt2dssa}
 S(t-\epsilon,x-\epsilon) + S(t-\epsilon,x+\epsilon) - S(t,x) - S(t - 2\epsilon,x) \geq 0.
\end{equation}
The inequality \eqref{alt2dssa}, upon expanding to second order in
$\epsilon$, immediately yields
\begin{equation}\label{2dlaplacian}
\eden_- \equiv \left( -\partial_t^2 + \partial_x^2 \right) S(t,x) \geq 0.
\end{equation}
Repeating the argument with the roles of $\lcgen^\pm$ reversed, we obtain $\eden_+ \geq 0$.

Note that the inequalities $\eden^\pm \geq 0$ can be saturated:
as is clear from the relation to the entropic c-function \cite{Casini:2004bw}, the entanglement densities $\eden_\pm$ are vanishing for the vacuum state of a  CFT. Furthermore, they also vanish whenever the EE can be computed in a CFT by a conformal transformation as in \cite{Calabrese:2004eu}, which includes, for example, the finite size system at zero temperature and the finite temperature system with an infinitely large size.

Physically, $\eden_\pm$ computes the entanglement between the two infinitesimally small light-like intervals $AC$ and $BD$ in Fig.~\ref{fig:2dSSA2}. Since both are directed in the opposite null directions, it is obvious that if the state is completely separated into the left and right-moving sector, the entanglement should be trivial. This explains why the entanglement density is vanishing for ground sates of 2d CFTs. On the other hand, for generic states, for example
a ground state of a non-conformal theory, we will find it is non-vanishing.

% Figure
%

%------------------------------------------------------------
\subsection{QFTs in $d=3$}
\label{subsec:3dssa}
%------------------------------------------------------------
%

The generalization to $d=3$ can be obtained following \cite{Casini:2012ei} by considering the iterated SSA inequality
\begin{equation}\label{hdssa}
\begin{split}
& \sum_{i} S \big(X_i \big) \geq S\left(\cup_i X_i\right) + S\left(\cup_{ij} \left( X_i \cap X_j \right) \right) \\
& \quad \quad + \; S \left(\cup_{ijk} \left( X_i \cap X_j \cap X_k\right) \right) + \dots + S\left(\cap_i X_i \right).
\end{split}
\end{equation}
We will work in a continuum limit, converting the sums to integrals on both sides of \eqref{hdssa}.

\begin{figure}[htbp]
\begin{center}
\includegraphics[width=2.3in]{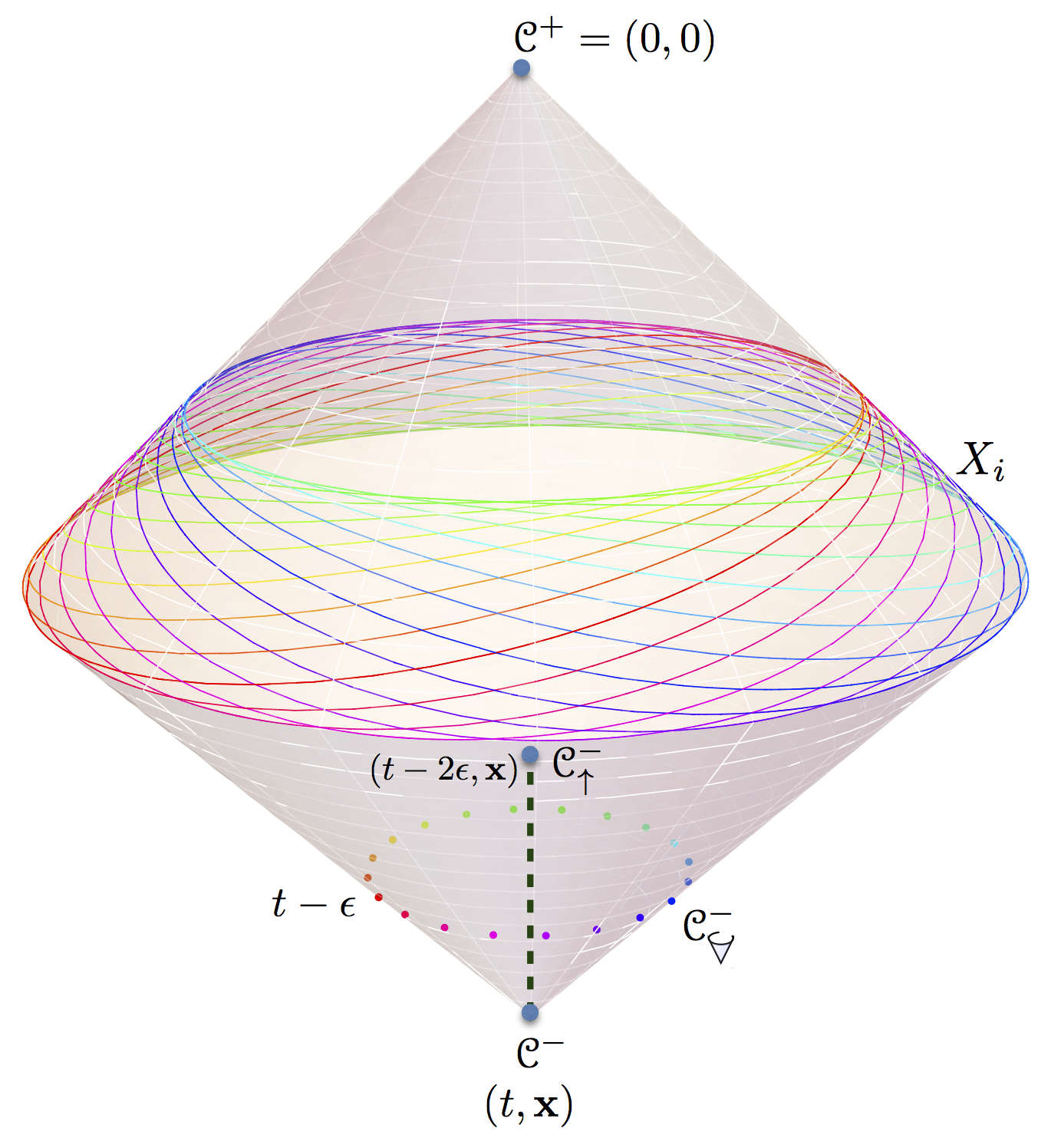}
\caption{Illustration of the set-up following \cite{Casini:2004bw} in $d\geq3$ with the same conventions as in
Fig.~\ref{fig:2dSSA2}. The regions $X_i$ in $d=3$ are obtained by considering the future light-cone from points distributed on the (dotted) circle, while their iterated intersections are obtained by considering the future light-cone from points on the (dashed) line-segment.}
\label{fig:3dSSA}
\end{center}
\end{figure}

We once again start with $\regA$ defined by $\lcgen^+ = (0,{\bf 0})$ and $\lcgen^- = (t,{\bf x})$. This corresponds to the choice of subsystem given by a round sphere. To apply SSA we consider translating $\lcgen^- \mapsto  \lcgen^-_{\scriptsize \cone}$ in the light-cone directions by a distance $\epsilon$, but this time respecting the rotation symmetry. This defines the subsystems $X_i$, described by ellipses  on
$\partial\domd{\regA}$. The loci of points composing $ \lcgen^-_{\scriptsize \cone}$ is a circle on $\partial J^+[\lcgen^-]$ at  time
$t-\epsilon$, as indicated in Fig.~\ref{fig:3dSSA}.

% Figure

To ascertain the unions of the iterated intersections on the r.h.s.\ of \eqref{hdssa} we make the following observation
\cite{Casini:2012ei}. Each term in the r.h.s.\ of \eqref{hdssa} generically leads to a curve which averages to a circular cross-section of the light-cone; in the present case we need cross-sections of $\partial J^-[\lcgen^+]$ at constant time. These can equivalently be obtained by translating $\lcgen^- \mapsto \lcgen^-_\uparrow$ in the temporal direction. With this in place we can examine the implications of SSA.

Consider first the contribution from the shift  $\lcgen^- \mapsto \lcgen^-_{\scriptsize \cone}$. Writing out the coordinates explicitly we find
\begin{align*}
\text{l.h.s.}_{\scriptsize{\eqref{hdssa}}}  = \left[1 - \epsilon\, \partial_t + \frac{\epsilon^2}{4} \left(\nabla^2_{{\bf x}} + 2\, \partial_t^2\right) \right] S(t,{\bf x})
+ {\cal O}(\epsilon^3) \,.
\end{align*}	
The r.h.s may be computed similarly, with the only additional complication being that we need to translate the measure from the circular cross-sections of $\partial J^-[\lcgen^+]$ onto the vertical segment along the map $\lcgen^- \mapsto \lcgen^-_\uparrow$. Accounting for this as in  \cite{Casini:2012ei} we find:
\begin{align*}
\text{r.h.s.}_{\scriptsize \eqref{hdssa}}
 =	  \left[1 -\epsilon\,\partial_t + \frac{\epsilon^2}{4}\, \left(3\, \partial_t^2 -\frac{2}{t}\, \partial_t \right)\right]
 	  S(t,{\bf x}) +\mathcal{O}(\epsilon^3) \,.
\end{align*}
 Combining the above two expressions we have the inequality resulting from SSA:
\begin{equation}
\eden_- \equiv \left[\Box+\frac{2}{t} \partial_t\right] S(t,{\bf x})  \geq 0 \, .
\end{equation}
Repeating the analysis about $\lcgen^+$ we can show $\eden_+ \geq 0$. This completes the derivation of \eqref{eq:edeneqn}.

Note that in boost invariant states (e.g., vacuum) where  $S_\regA$ is a function of
proper length $\ell  = \sqrt{t^2 - ||{\bf x}||^2}$, \eqref{eq:edeneqn} simply reduces to  \cite{Casini:2004bw,Casini:2012ei}:
\begin{equation}
 \ell\, S''(\ell) - (d-3) \,S'(\ell) \leq 0\,.
 \label{leqn}
\end{equation}
We have however managed to convert this to a local statement for regions $\regA$ which are naturally generated by intersecting  light-cones from two points $\lcgen^\pm$.
Although we have written the expressions \eqref{eq:edeneqn} and \eqref{leqn} in a manner which suggests an obvious generalization to higher $d$, there are some subtleties with this interpretation, which we revisit in \S\ref{sec:discuss}.

%~~~~~~~~~~~~~~~~~~~~~~~~~~~~~~~~~~~~~~~~~~~~~~~
\section{Holographic entanglement density}
\label{sec:gtew}
%~~~~~~~~~~~~~~~~~~~~~~~~~~~~~~~~~~~~~~~~~~~~~~

Having understood the basic constraint on the entanglement density, let us now consider the holographic context, employing the AdS$_3/$CFT$_2$ duality. We focus on linear perturbations around the pure AdS$_3$ solution, corresponding to small excitations around the vacuum. In the bulk gravity theory, we consider  Einstein gravity coupled to arbitrary matter fields, with the energy-momentum tensor $T_{\mu\nu}$ given by the Einstein's equation
\begin{equation}
E_{\mu\nu} = R_{\mu\nu}-\f{1}{2}Rg_{\mu\nu}+\Lambda \,g_{\mu\nu} = 8\pi G_N\, T_{\mu\nu}.
\label{eq:eins}
\end{equation}	

It is now convenient to work directly with the end-points of $\regA$, whose null coordinates are
$(\uL,\vL)$ and $(\uR,\vR)$ respectively in
${\mathbb R}^{1,1}$. In terms of these, the two entanglement densities are given by:
\begin{equation}
\eden_+=-\f{\de}{\de \uR}\f{\de}{\de \vL}\Delta S_{\regA},
\quad \hat{n}_-=-\f{\de}{\de \uL}\f{\de}{\de \vR}\Delta S_{\regA} \,.
\label{eq:edenRel}
\end{equation}	
Note that we define the density in terms of $\Delta S_{\regA} = S_\regA^{\rho_\Sigma} - S_\regA^{vac}$ which measures the entanglement of the excited state $\rho_\Sigma$ relative to the vacuum. It is crucial here that $\eden_\pm$ vanishes in the vacuum state, for while SSA holds for any state of the CFT, it is no longer true that $\Delta S_\regA$ satisfies SSA.\footnote{ It is easy to verify this statement explicitly say by considering $\rho_\Sigma$ to be the thermal state.} With this understanding we can replace $S_\regA \to \Delta S_\regA$ and still maintain the sign-definiteness of entanglement densities $\hat{n}_{\pm}$  defined in \eqref{eq:edenRel}.

We now evaluate $\Delta S_{\regA}$ by analyzing the holographic entanglement entropy in the perturbed geometry around  pure AdS$_3$ described by the (gauge fixed) metric:
\begin{equation}
ds^2=\frac{dz^2-du\,dv}{z^2}+h_{ab}(u,v,z) \, dx^a\, dx^b\,,
\end{equation}	
where $h_{ab}$ captures the perturbation (Latin indices refer to the boundary). For linear order changes of holographic entanglement entropy, we can work with the original geodesic in AdS$_3$ (parameterized by $\xi$) which connects the endpoints of $\regA$:
\begin{equation*}
(u,v,z)= (\um+   \ud \sin\xi,\
\vm+\vd \sin\xi,\ \s{|\ud\vd|}  \cos\xi)\,,
\end{equation*}	
where $\{U,\ud\} = \frac{1}{2}(\uR\pm\uL)$ and
$\{V,\vd\} =\frac{1}{2}\, (\vR\pm\vL)$ give the mid-point and separation between the end-points of $\regA$.

The first-order perturbation of $\Delta S_{\regA}$ is given by
\begin{equation}
\Delta S_A=\f{1}{8\,G_N}\int^{\f{\pi}{2}}_{-\f{\pi}{2}} d\xi \f{\gamma^{(1)}(\xi)}{\s{\gamma^{(0)}(\xi)}}\,,
\end{equation}	
where $\gamma^{(0)}$ and $\gamma^{(1)}$ are induced metric ($\gamma_{\xi\xi}$) at leading and first sub-leading orders, i.e.,
\begin{align*}
\gamma^{(0)}(\xi) &= \f{1}{\cos^2\xi}\,, \nonumber \\
\gamma^{(1)}(\xi) &= \cos^2\xi   \left(h_{uu}\,\ud^2+h_{vv} \,\vd^2 + 2\,h_{uv}\, \ud \,\vd \right)\,.
\end{align*}

After some algebra we arrive at the following simple relations:
\begin{align}
\eden_{\pm}&=
	\f{1}{4\,G_N\, |\ud\vd| } \, \int^{\f{\pi}{2}}_{-\f{\pi}{2}}
	d\xi
	\s{\gamma^{(0)}} \left(N_{(\pm)}^\mu N_{(\pm)}^\nu E_{\mu\nu}\right) \nonumber \\
&=
	\frac{2\pi}{|\ud\vd|}\, \int^{\f{\pi}{2}}_{-\f{\pi}{2}}
	d\xi \s{\gamma^{(0)}}
	\left(N_{(\pm)}^\mu N_{(\pm)}^\nu T_{\mu\nu}\right)
	\geq 0 \,,
\label{einw}
\end{align}
where $E_{\mu\nu}$ is the l.h.s.\ of the Einstein's equation \eqref{eq:eins}.
The vectors  $N_{(\pm)}^\mu$  are the two independent null normals to the extremal surface
$\extr$ in AdS$_3$,
\begin{align*}
 N_{(\pm)}^\mu =
\bigg\{
%\eta\,
 \frac{\ud\, \cos^3\xi}{(\sin\xi \mp1)},
\frac{\vd\,\cos^3\xi}{(\sin\xi \pm 1)} ,-\sqrt{|\ud\vd|}
 \cos^2\xi\bigg\} \,.
\end{align*}

Firstly, we note from (\ref{einw}) that the positivity of entanglement density is correlated with null energy condition.
While we have established the  above result explicitly only for linear deviations away from the vacuum, the fact that
$\eden_\pm$ vanishes in  vacuum, and its positive semi-definiteness from SSA for any excited state, makes it natural for us to
conjecture that  the relation
\begin{equation}
\eden_{\pm}
=\f{1}{8\,G_N}\int_{\extr} d\xi \s{\gamma_{\xi\xi}}\left(N_{(\pm)}^\mu N_{(\pm)}^\nu \,E_{\mu\nu}\right)
\geq 0
\label{eq:nullE3}
\end{equation}	
holds for any asymptotically AdS$_3$ backgrounds, with $\extr$ being the extremal surface
(spacelike geodesic parameterized $\xi$) which holographically encodes $S_{\regA}$. We leave a more complete exploration of this relation for  the future.

It is interesting to note that for normalizable states of pure gravity in AdS$_{3}$, the entanglement density always vanishes. This is consistent with our earlier observation that entanglement density is vanishing for any state obtained by conformal transformations of ground states in 2d CFTs. Indeed, solutions in the pure AdS$_3$ gravity can be obtained by bulk diffeomorphisms corresponding to boundary conformal transformations \cite{Skenderis:1999nb}.

%~~~~~~~~~~~~~~~~~~~~~~~~~~~~~~~~~~~~~~~~~~~~~~~
\section{Discussion}
\label{sec:discuss}
%~~~~~~~~~~~~~~~~~~~~~~~~~~~~~~~~~~~~~~~~~~~~~~

In this paper we have introduced a new quantity, the entanglement density $\eden$ for relativistic field theories, and argued that it provides a useful encoding of certain aspects of gravitational dynamics via holography. We have directly argued for its positivity using the SSA property of EE in  2d and 3d field theories. More generally, we see from our explicit analysis that the positivity of $\eden$ and the gravitational null energy condition go hand in hand.
 At the same time, we anticipate (\ref{eq:nullE3}) to be  of fundamental importance, since it geometrically encodes the SSA and captures second order variations of holographic entanglement entropy.

While our holographic analysis was carried out for linearized fluctuations around AdS$_3$, we anticipate that \eqref{eq:nullE3}
holds at the  non-linear level. In fact, it is tempting to conjecture a more general statement valid in any dimension:
SSA implies that the entanglement density $\eden \geq 0$ for any state of a QFT with $\eden^{vac} =0$.
Furthermore, translating the description of $\eden$ into holography one finds that \eqref{eq:enngr} holds for any deformation away from pure AdS in arbitrary spacetime dimensions. To wit,
\begin{align}
&\text{SSA}  \;\Longrightarrow \;\eden_{\pm} \geq 0 \,, \;\;\eden^{vac}_\pm = 0 \,,
\nonumber \\
& \Longrightarrow
\int_{\extr} {\bm \epsilon} \; N_{(\pm)}^\mu N_{(\pm)}^\nu \,E_{\mu\nu}
\geq 0\,,
\end{align}	

One could try to follow the logic of \S\ref{subsec:3dssa} to arrive at the conclusions above, by considering variations of  the past tip of $\domd{\regA}$ (cf., Fig.~\ref{fig:3dSSA} with each point replaced by  ${\bf S}^{d-3}$). However, this attempt runs afoul of sub-leading divergences in the entanglement entropy from the r.h.s.\ of \eqref{hdssa} as explained in \cite{Casini:2012ei}. It is nevertheless interesting to contemplate whether the entanglement density can be used to provide further insight into c and F-theorems and generalizations thereof.

Nevertheless we may draw the following analogy based on the conjecture above: the statement of SSA is
reminiscent of the second law of thermodynamics since it asserts convexity of entanglement (but under region variation as opposed to time variation). We are arguing that this guarantees positivity of the entanglement density. Via holography, generic deformations
about the CFT vacuum (equilibrium) then increase the `cosmological Einstein tensor' $E_{\mu\nu} $ when suitably averaged over the extremal surface. In essence, this quantity codifies a version of gravitational second law for entanglement density. Indeed,
in the `long-wavelength'  (hydrodynamic) regime, one may capture the thermal entropy production via the entanglement density
by taking $\regA$ to be suitably large.

%%%%%%%%%%%%%%%%%%%%%%%%%%%%%%%%%%%%%%%%%%%%%%%%%%%%%
\begin{acknowledgements}
%%%%%%%%%%%%%%%%%%%%%%%%%%%%%%%%%%%%%%%%%%%%%%%%%%%%%

It is a pleasure to thank Shamik Banerjee, Horacio Casini, Matt Headrick,  Juan Maldacena, Emil Martinec,
Tatsuma Nishioka, and Masahiro Nozaki for discussions.

VH, MR would like to thank the IAS, Princeton, YITP, Kyoto, U. Amsterdam and Aspen Center for  Physics (supported by the National Science Foundation under Grant 1066293) for hospitality during the course of this project.

JB is supported by the STFC Consolidated Grant ST/L000407/1.  VH and MR were supported in part by the Ambrose Monell foundation, by the FQXi grant ``Measures of Holographic Information'' (FQXi-RFP3-1334), by the STFC Consolidated Grants ST/J000426/1 and  ST/L000407/1, and by the NSF grant under Grant No. PHY-1066293.
MR also acknowledges support from the ERC Consolidator Grant Agreement ERC-2013-CoG-615443: SPiN. TT is supported by JSPS Grant-in-Aid for Scientific Research (B) No.25287058 and by JSPS Grant-in-Aid for Challenging
Exploratory Research No.24654057. TT is also supported by World Premier International Research Center Initiative (WPI Initiative) from the Japan Ministry of Education, Culture, Sports, Science and Technology (MEXT).

\end{acknowledgements}

%%%%%%%%%%%%%%%%%%%%%%%%%%%%%%%%%%%%%%%%%%%%%%%

% \bibliographystyle{apsrev}
% \bibliography{edensity}

\end{document}